\documentclass[prd,aps,nofootinbib,floatfix,11 pt]{revtex4}
\usepackage{amssymb}
\usepackage{amsmath,graphicx,color,epsfig}
\begin{document}

\title{Dynamical Symmetries of Dirac Hamiltonian}
\author{Riazuddin}
\email{riazuddin@ncp.edu.pk}
\affiliation{National Centre for Physics, Quaid-i-Azam University Campus, Islamabad 45320, Pakistan}
\date{\today}

\begin{abstract}
Several dynamical symmetries of the Dirac Hamiltonian are reviewed
in a systematic manner and the conditions under which such
symmetries hold. These include relativistic spin and orbital angular
momentum symmetries, $SO(4)\times SU_{\sigma }(2)$ symmetry for the
Dirac Hydrogen atom, $SU(3)\times SU_{\sigma }(2)$ symmetry for the
relativistic simple harmonic oscillator. The energy spectrum in each
case is calculated from group-theoretic considerations.
\end{abstract}

\maketitle

\section{Introduction}

The $1/r$ potential of force provides the underlying dynamics for
the Kepler problem in classical mechanics, hydrogen atom in quantum
mechanics and one gluon exchange potential in perturbative quantum
electrodynamics (PQCD). This and simple harmonic oscillator are one
of the few physical systems that we know how to solve. Further most
of more complicated systems can be studied by representing them as a
collection of harmonic oscillators with different frequencies and
amplitudes. Not only the above two systems are solvable exactly both
in classical and quantum mechanics, they also provide the realistic
models to study dynamical symmetries as distinct from geometrical
symmetries. Symmetries do play an important role in the progress of
physics. Once they are known for a physical system, many of the
properties of that system can be established in general terms
without actually solving for the underlying dynamics.

As stated above symmetries are of two types: (i) Geometric, the well
known example of which are space-time symmetries, e.g. rotation (ii)
Dynamical, where the underlying dynamics manifest some symmetry.
This is well illustrated both by the hydrogen atom and harmonic
oscillator, which are rotationally invariant but also have a well
known feature that their orbits of motion close on themselves in
classical mechanics. This shows that in addition to angular momentum
(conservation which is due to rotational invariance), there are
additional constants of motion. For the hydrogen atom this constant
of motion, as is well known, is provided by the Rung-Lenz's vector
\cite{01}.  For the harmonic oscillator this is provided by
quadrupole moment second rank tensors \cite{moment}.

For the hydrogen atom in Schrodinger theory, energy levels are given
by $E_{n}=-\frac{m\alpha}{2n^{2}}$, where $n$ is the principle
quantum number, and $\alpha$ is the fine structure constant. For a
given $l$, $n\geq l+1$; $l=0,1,2,....n$. The $(2l+1)$ degeneracy
with respect to magnetic quantum number is due to rotational
symmetry of the potential $-\frac{\alpha}{r}$. For one energy value
$E,$ there are $\sum\limits_{l=0}^{n-1}\left( 2l+1\right) =n^{2}$
different possible eigen-functions; such a degeneracy occurs only
for $1/r$ potential of force. Usually a degeneracy is associated
with a symmetry and it is known \cite{01} that in this case there is
an external symmetry $\left[ \vec{R},H\right] =0,$with $H=\frac{\vec{p}^{2}}{2m}%
-\frac{\alpha}{r}$, and $\vec{R}=\frac{1}{2m}(\vec{p}\times \vec{L}-\vec{L}%
\times \vec{p})-\frac{\alpha}{r}\vec{r}$. The operator $\vec{R}$
given above in Quantum Mechanics follows from the Lenz's vector in
the classical Kepler problem by the correspondence principle. The
orbital angular momentum $\vec{L}$ and
$\vec{K}=\sqrt{\frac{-m}{2H}}\vec{R}$ ($\vec{K}$ is hermitian when
acting on eigenstates of $H$ with-negative
energy eigenvalues- the ones in which we are interested; $\sqrt{\frac{-m}{2H}%
}$ commutes with $\vec{R}$\ and $\vec{L}$) generate $SO(4)$ algebra, which
is isomorphic to $SU_{M}(2)$ $\times SU_{N}(2)$ generated by $\vec{M}=\frac{%
\vec{L}+\vec{K}}{2}$, $\vec{N}=\frac{\vec{L}-\vec{K}}{2}$. This
symmetry leads to $n^{2}$ degeneracy \cite{01}.

Similarly for the non- relativistic harmonic oscillator, the energy
spectrum shows degeneracies in addition to those which arise due to
rotational invariance. As is well known, the energy spectrum
$E_{N}=\frac{1}{2}\hbar w(2N+3)$ depends on the oscillator quantum
number $N=2n+l,$ where $n\geq 0$ is the radial quantum number and
$l$ is the orbital angular momentum. Thus all states with
$l=N,N-2,....0$ or $1$ have the same energy. These degeneracies are
produced by an $SU(3)$ dynamical symmetry \cite{02}.

Obviously the above symmetries are broken in the relativistic
quantum mechanics, where a spin 1/2 particle satisfies the Dirac
equation. As is well known this is because the spin-orbit coupling
leads to splitting of the energy levels [see below]. Since spin is
purely a quantum mechanical concept with no classical analogue, the
guidance from the corresponding principle is not there to obtain
dynamical symmetries for the Dirac Hamiltonian. As such it is highly
non trivial to discuss under which conditions the larger symmetries
are obtained in the Dirac Hamiltonian. This has been answered for
harmonic oscillator in \cite{03} and following the method of
\cite{03}, for the Hydrogen atom in \cite{04}. The purpose of this
paper is to review the previous work mentioned above and to discuss
the symmetries of the Dirac Hamiltonian in a systematic way, using
the Dirac algebra and for harmonic Oscillator $SU(3)$ algebra
generated in the Gell-Mann basis \cite{05} and point out which
physical system might have such symmetries. In fact we first show
that the harmonic oscillator provides a simple realization of
$SU(3)$, which has been used in particle and nuclear physics. The
energy spectrum in each case has also been calculated from
group-theoretic considerations.

\section{Dynamical spin and orbital angular momentum symmetry for the Dirac Hamiltonian}

The Dirac Hamiltonian is given by
\begin{equation}
H=\vec{\alpha}\cdot \vec{p}+V_{V}\text{ }(\vec{r})+\beta m
\label{00}
\end{equation}%
where for the Hydrogen atom $V_{V}$ $(\vec{r})=-\frac{\alpha}{r}$
and is the time component of electromagnetic potential $A^{\mu }$
$(\vec{r}),$ $(\mu =0,1,2,3).$ The hydrogen atom in Dirac theory is
exactly solvable \cite{06,dt1} and energy levels are given by

\begin{equation*}
E_{nj}=m[1+\frac{\alpha^{2}}{(n-\delta_{j})^{2}}]^{-1/2}
\end{equation*}
where
\begin{equation}
\delta_{j}=(j+\frac{1}{2})-[(j+\frac{1}{2})^{2}-\alpha^{2}]^{1/2}
\label{01}
\end{equation}
and $j=0,1/2,1...$. The above expression manifests the fine
structure and spin-orbital splitting. The states with same $n$ but
different $j$ e.g. $2p_{1/2}$ and $2p_{3/2}$ are now split and the
splitting is in agreement with experiment , a great triumph of the
Dirac theory. The states with same $n$ and $j$ e.g. $2s_{1/2}$ and
$2p_{1/2}$ are still degenerate, the so called Lamb shift which
needs radiative quantum corrections provided by quantum
electrodynamics (QED). The SO(4) symmetry for the hydrogen atom in
Schrodinger theory is thus broken in relativistic theory. Usually in
the non-relativistic limit i.e to order $(p^{2}/m^{2})$, the
relativistic theory reduces to non-relativistic theory but in this
case the non-relativistic theory has a higher symmetry, namely,
SO(4) which is not maintained by the relativistic theory. The
question then arises under what circumstances for hydrogen-like
system (by this we mean systems with $1/r$ force potential), the
SO(4) symmetry is restored for the Dirac Hamiltonian, although it
will be contrary to experimental observations for the hydrogen atom.
Then why should even this question be considered? This is because
there are other physical systems which show such an approximate
symmetry as we will discuss shortly. To answer the above question,
let us introduce a Lorentz scalar potential $V_{S}$ $(\vec{r}),$
then the Dirac Hamiltonian becomes
\begin{equation}
H=\vec{\alpha}\cdot \vec{p}+V_{V}\text{ }(\vec{r})+\beta (V_{S}\text{ }(\vec{%
r})+m)  \label{02}
\end{equation}%
and the corresponding Dirac equation is

\begin{equation}
\lbrack i\gamma ^{0}(\partial _{0}+iV_{v}(r))+i\gamma ^{i}\partial
_{i}-m-V_{s}(r)]\Psi =0.
\end{equation}
 where $\gamma^{i}=\beta\alpha^{i},
\gamma^{0}=\beta$

If we multiply on the left by
\begin{equation*}
\lbrack i\gamma ^{0}(\partial _{0}+iV_{V}(r))+i\gamma ^{j}\partial
_{j}+m+V_{S}(r)].
\end{equation*}
Then for stationary states ($\frac{\partial }{\partial t}\rightarrow
iE$) , the Dirac equation becomes
\begin{equation}
\lbrack \nabla ^{2}+V_{V}^{2}-V_{S}^{2}-2EV_{V}-2mV_{S}+i\gamma
^{0}\gamma ^{i}[\partial _{i},V_{V}]-i\gamma ^{i}[\partial
_{i},V_{S}]+(E^{2}-m^{2})]\Psi =0.
\end{equation}
where
\begin{eqnarray}
\lbrack \partial _{i},V_{V}] &=&\frac{\partial V}{\partial
x^{i}}, \nonumber \\
 \lbrack \partial _{j},V_{S}]
&=&\frac{\partial V}{\partial x^{j}}. \label{03a}
\end{eqnarray}

Thus if
\begin{eqnarray}
V_{V}(\overrightarrow{r})=V(\overrightarrow{r})+U_{V} \nonumber \\
V_{S}(\overrightarrow{r})=V(\overrightarrow{r})+U_{S}\label{03b}
\end{eqnarray}
where $U_{V}$ and $U_{S}$ are constants, then since in the
non-relativistic limit
$\gamma^{0}\rightarrow1+O(\frac{p^{2}}{m^{2}}),
\gamma^{i}\rightarrow1+O(\frac{p^{i}}{m}), E\rightarrow\epsilon +m$,
the terms of order $\frac{|p|}{m}$ cancel out in Eq. (5). As a
consequence absorbing the constants $U_{V}$ and $U_{S}$ with
re-definition of $E$ and $m$, the above equation reduces to the
Schrodinger equation, which as already seen has $SO(4)$ symmetry, if
$V$ is spherically symmetric.

Indeed it has been observed that the Dirac Hamiltonian (3) is
invariant under a spin symmetry \cite{dt2}, $\left[ H,\vec{S}\right]
=0,$ provided that conditions (6) or (7) are satisfied. Here the
generators $\vec{S}$ form the spin $SU(2)$ algebra and are given by
\begin{equation}
\vec{S}=\left(
\begin{array}{cc}
\vec{s} & 0 \\
0\text{ } & u_{p}\text{ }\vec{s}\text{ }u_{p}%
\end{array}%
\right)  \label{03}
\end{equation}%
where $\vec{s}=\vec{\sigma}/2$ are usual spin generators and $u_{p}=\frac{%
\vec{\sigma}\cdot \vec{p}}{p}$ is the helicity unitary operator. It is easy
to check that
\begin{equation}
\left[ S_{i},S_{j}\right] =i\text{ }\epsilon _{ijk}\text{ }S_{k}  \label{04}
\end{equation}%
In the Pauli representation of Dirac matrices,%
\begin{equation}
\beta =\left(
\begin{array}{cc}
1 & 0 \\
0 & -1%
\end{array}%
\right) ,\ \ \ \vec{\alpha}=\left(
\begin{array}{cc}
0 & \vec{\sigma} \\
\vec{\sigma} & 0%
\end{array}%
\right) ,\ \ \gamma ^{5}=-i\alpha ^{1}\alpha ^{2}\alpha ^{3}=\left(
\begin{array}{cc}
0 & 1 \\
1 & 0%
\end{array}%
\right)  \label{05}
\end{equation}%
By introducing $\vec{\Sigma}=\left(
\begin{array}{cc}
\vec{\sigma} & 0 \\
0 & \vec{\sigma}%
\end{array}%
\right) ,$ one can write the Dirac Hamiltonian (3) and the spin
operator (8) as
\begin{equation}
H=\gamma ^{5}\beta \text{ }\vec{\Sigma}\cdot \vec{p}+V_{V}\text{ }(\vec{r}%
)+\beta (V_{S}\text{ }(\vec{r})+m)  \label{06}
\end{equation}%
\begin{equation}
\vec{S}=\frac{1}{2}\left[ \beta \vec{\Sigma}+\left( 1-\beta \right) \vec{%
\Sigma}\cdot \vec{p}\text{ }\vec{p}\frac{1}{p^{2}}\right]   \label{07}
\end{equation}%
Then, since $\left[ \beta ,1-\beta \right] =0,$ \ \ \ $\left[ \beta ,\vec{%
\Sigma}\right] =0,$ \ \ $\left[ \gamma ^{5},\vec{\Sigma}\right] =0,$\ $\left[
\gamma ^{5},\beta \right] _{+}=0$ and $\left[ \vec{\Sigma}\cdot \vec{p},\vec{%
\Sigma}\cdot \vec{p}\text{ }p^{i}\right] =0$,
\begin{eqnarray}
\left[ H,\vec{S}\right]  &=&\frac{1}{4}\left[ \gamma ^{5},\beta \right] %
\left[ \vec{\Sigma}\cdot \vec{p}\text{ },\text{ }\Sigma ^{i}\right] _{+}%
\text{ +}\frac{1}{4}\left[ \gamma ^{5},(1-\beta )\right] \left[ \vec{\Sigma}%
\cdot \vec{p},\frac{\vec{\Sigma}\cdot \vec{p}\text{ }p^{i}}{p^{2}}\right]
_{+}  \notag \\
&&+\frac{\left( 1-\beta \right) }{2}\left[ V_{V}\text{ }(\vec{r}),\left(
1-\beta \right) \frac{\vec{\Sigma}\cdot \vec{p}\text{ }\vec{p}}{p^{2}}\right]
+\frac{\beta \left( 1-\beta \right) }{2}\left[ V_{S}\text{ }(\vec{r}),\frac{%
\vec{\Sigma}\cdot \vec{p}\text{ }}{p^{2}}p^{i}\right]   \notag \\
&=&\frac{1}{2}\gamma ^{5}\beta \text{ }2\text{ }\delta ^{ij}p^{j}-\frac{1}{2}%
\gamma ^{5}\beta \text{ }2\text{ }p^{i}\text{\ }+\frac{\left( 1-\beta
\right) }{2}\Sigma ^{j}\text{ }\left[ \left( V_{V}\text{ }(\vec{r})-V_{S}%
\text{ }(\vec{r})\right) ,\frac{p^{j}p^{i}}{p^{2}}\right]   \label{08}
\end{eqnarray}%
Thus $\left[ H,\vec{S}\right] =0$\ if $\partial ^{i}V_{V}$ $(\vec{r}
)=\partial ^{i}V_{S}$ $(\vec{r})$ or $V_{V}$ $(\vec{r})=V_{S}$
$(\vec{r})+U$ i.e. if the conditions (6) or (7) are satisfied.
Further for spherically symmetric potentials, $V_{V}$
$(\vec{r})=V_{V}$ $ (r),$ $\ V_{S}$ $(\vec{r})=V_{S}$ $(r),$ the
Dirac Hamiltonian has an additional invariant algebra \cite{dt2},
$\left[ H,\vec{L}\right] =0$ where
\begin{equation}
\vec{L}=\left(
\begin{array}{cc}
\vec{l} & 0 \\
0\text{ } & u_{p}\text{ }\vec{l}\text{ }u_{p}%
\end{array}
\right)   \label{09}
\end{equation}
and $\vec{l}=\vec{r}\times \vec{p}$ is the orbital angular momentum.
One can write (14) as
\begin{equation}
\vec{L}=\vec{l}+\frac{\left( 1-\beta \right) }{2}\left[
\vec{\Sigma}-\vec{ \Sigma}\cdot \vec{p}\text{ }\vec{p}/p^{2}\right]
\label{010}
\end{equation}
Then $\left[ V_{V}\text{ }(r),\vec{l}\right] =0,$ $\left[
V_{S}\text{ }
(r),\vec{l}\right] =0$ and it follows as above that%
\begin{equation}
\left[ H,\vec{L}\right] =0,\text{ \ \ \ }\left[ L_{i},L_{j}\right] =i\text{ }%
\epsilon _{ijk}\text{ }L_{k}  \label{011}
\end{equation}%
Thus $\vec{S}$ and $\vec{L}$ are separately constants of motion while $\vec{s%
}$ and $\vec{l}$ are not, but $\vec{s}$ $+$ $\vec{l}$ is.

Finally the Dirac Hamiltonian, which has the above relativistic
dynamical spin and orbital angular momentum symmetries, is
\begin{equation}
H=\gamma ^{5}\vec{\Sigma}\cdot \vec{p}+(1+\beta )V(r)+\beta m
\label{012}
\end{equation}%
where constant $U$ can be absorbed in the mass term $m$. We note that
\begin{equation}
(H+m)=\gamma ^{5}\vec{\Sigma}\cdot \vec{p}+(1+\beta )(V+m)
\label{012a}
\end{equation}
\begin{equation}
H^{2}-m^{2}=\vec{p}^{2}+2(1+\beta)V(m+V)+\gamma
^{5}\left[(1-\beta)V\vec{\Sigma}\cdot \vec{p}+(1+\beta
)V\vec{\Sigma}\cdot \vec{p}\right]\label{012b}
\end{equation}

There is another symmetry the so called relativistic pseudo spin and
orbital angular momentum symmetry \cite{dt2}, which is obtained from
the one considered above by making the $\gamma_{5}$ transformation
and $m\rightarrow -m$

\begin{equation*}
H\rightarrow \widetilde{H}=\gamma^{5} H \gamma^{5}=\gamma^{5}
\vec{\Sigma}\cdot \vec{p}+(1-\beta)V(r)+\beta m
\end{equation*}

\begin{equation*}
 \vec{S} \rightarrow \widetilde{\overrightarrow{S}}=\gamma^{5}S \gamma^{5}
=\frac{1}{2}[- \beta \vec{\Sigma}+\left( 1+\beta
\right)\frac{\vec{\Sigma} \cdot
 \vec{p}\vec{p}}{p^{2}}]
\end{equation*}

\begin{equation*}
 \vec{L} \rightarrow \widetilde{\overrightarrow{L}}=\vec{l}+\frac{1+\beta}{2}[\vec{\Sigma}-\frac{\vec{\Sigma} \cdot
 \vec{p}\vec{p}}{p^{2}}]
\end{equation*}

\begin{equation}
[\widetilde{H},\widetilde{ \overrightarrow{S}}]=0,
[\widetilde{H},\widetilde{ \overrightarrow{L}}]=0
\end{equation}

The first of Eqs. (20) implies $V_{V}=-V_{S}+$ constant for
pseudo-spin symmetry. The pseudo-spin degeneracies have been
observed in nuclei\cite {07}. In fact relativistic mean field
representations of the nuclear potential do have this property \cite
{08} namely $V_{V}\simeq-V_{S}$.

We now consider the question wether the property
$V_{V}(\overrightarrow{r})=V_{S}(\overrightarrow{r})+U$ arises in
QCD. In the $(Q\bar q)$ or $(q\bar Q)$ bound states meson
spectroscopy, where $Q$ is a heavy quark $c$ or $b$ and $q$ is light
quark $u,d,$ or $s$, the spin-orbital splitting is seen to be
suppressed experimentally \cite {010}. In a $(Q\bar q)$ system,
where $Q$ is considered to be infinitely heavy, the spin
$\overrightarrow{S_{Q}}$ of the heavy quark is decoupled, it is then
natural to combine the angular momentum of light degrees of freedom
$\overrightarrow{j}=\overrightarrow{L}+\overrightarrow{S_{q}}$ with
$\overrightarrow{S_{Q}}$ to give
$\overrightarrow{J}=\overrightarrow{j}+\overrightarrow{S_{Q}}$ for
the bound $(Q\bar q)$ system. Thus for the $p$ states we get two
multiplets, one with $j=3/2$ and other with $j=1/2$ which for $Q=c$,
and $q=u $ or $d$ are [and similar one for $B$ meson when $Q=b$]
\begin{eqnarray*}
l &=&0\text{ \ \ \ \ [}D^{\ast }(1^{-}),D(0^{-})\text{]}_{j=\frac{1}{2}}, \\
l &=&1\text{ }\ \ \ \ \text{[}D_{2}^{\ast
}(2^{+}),D_{1}(1^{+})\text{]}_{j=
\frac{3}{2}}, \\
&&\text{ \ \ \ \ \ \ \ }and\text{ \ [D}_{1}^{\ast
}(1^{+}),D_{0}(0^{+})\text{ ]}_{j=\frac{1}{2}},
\end{eqnarray*}

The splitting between $j=\frac{3}{2}$ and $j=\frac{1}{2}$ multiplets
is due to spin-orbit coupling $\overrightarrow{L}\cdot
\overrightarrow{S_{q}}$ (as in hydrogen atom) while the hyperfine
splitting between the two members of each multiplet arises from the
Fermi-contact term $\overrightarrow{S_{q}}\cdot
\overrightarrow{S_{Q}}$, the spin-orbit coupling term
$(\overrightarrow{S_{q}}+ \overrightarrow{S_{Q}})\cdot
\overrightarrow{L}$ and the tensor terms. The splitting [12], for
$D$ mesons, between $D_{1}^{\ast }(1^{+}):2422.3\pm 0.6MeV$ and
$D_{2}^{\ast
}(2^{+}):2462.8\pm 1.0MeV$ is $40MeV$; for $B$ mesons, between $%
B_{1}(1^{+}):5723.4\pm 2.0$ and $B_{2}^{\ast }$($2^{+}):5743.9\pm 5$ is $%
20MeV$; for $B_{s_{1}}(1^{+}):5829.4\pm 0.7$ and
$B_{s_{2}}(2^{+}):5839.7\pm 0.6$ is $10MeV$. We notice that the spin
orbit splittings seem to be suppressed.  A measure of this
suppression \cite {09} is the parameter
\[
r=\frac{p_{3/2}-p_{1/2}}{(4p_{3/2}+2p_{1/2})/6-s_{1/2}}
\]
which for the experimental data shown above is of order $0.07$ both
for $D$ and $B$ mesons. We also note that suppression generally
increases with the increasing mass of $Q$.
 It has been suggested \cite {09} that the dynamics
necessary for the approximate relativistic spin symmetry discussed
above in the heavy-light quark system may be possible in QCD.

\section{$SO(4)\times SU_{\sigma }(2)$ Symmetry for a Coulomb
like Symmetry Potential}

 In this section we discuss the generators of $SO(4)$ symmetry, which the Dirac Hamiltonian (3) possess. Since in the
non-relativistic limit $\beta \rightarrow 1+O(p^{2}/m^{2})$ and
$\gamma ^{5}\rightarrow O(p/m),$ the natural generalization of the
Lenz's
vector in the Schrodinger theory, for the relativistic case is%
\begin{equation*}
2m\vec{R}\rightarrow 2m\vec{\Gamma}
\end{equation*}%
where
\begin{equation}
2m\vec{\Gamma}=\left( 1+\beta \right) f(r)\vec{r}+%
\vec{\Lambda}+\left[ \left( 1+\beta \right) \gamma _{5}g(r)%
\vec{r}\vec{\Sigma}\cdot \vec{p}+\left( 1-\beta \right)
\gamma _{5}\vec{\Sigma}\cdot \vec{p}g(r)\vec{r}%
\right]   \label{013}
\end{equation}%
where now
\begin{equation}
\vec{\Lambda}=\vec{p}\times \vec{L}-\vec{L}\times \vec{p}  \label{014}
\end{equation}%
$\vec{L}$ is the relativistic orbital angular momentum defined in
Eq. (14). Since $H$ involves $\gamma ^{5}\vec{\Sigma}\cdot \vec{p}$,
therefore $\vec{\Gamma}$ should involve such a term and the second
term in square
brackets appears to make the operator hermitian. The functions $f(r)$ and $%
g(r)$ are to be determined from
\begin{equation}
\left[ H,\vec{\Gamma}\right] =0  \label{015}
\end{equation}%
Now $\vec{L}$\ commutes with $H$ and also with $V$ and $\beta m$\ \ and
therefore it also commutes with $\gamma ^{5}\vec{\Sigma}\cdot \vec{p}$, it
follows that $\left[ \gamma ^{5}\vec{\Sigma}\cdot \vec{p},\vec{\Lambda}%
\right] =0$\ since $\vec{p}$\ commutes with $\vec{\Sigma}\cdot \vec{p}$.
Thus using $(\vec{\Sigma}\cdot \vec{p})^{2}=p^{2},$%
\begin{eqnarray}
2m\left[ H,\vec{\Gamma}\right]  &=&\gamma ^{5}\left[ \vec{\Sigma}\cdot \vec{p%
},\text{ }f(r)\text{ }\vec{r}-2(V+M)g(r)\text{ }\vec{r}\right] +\gamma
^{5}\beta \left[ \vec{\Sigma}\cdot \vec{p},\text{ }f(r)\text{ }\vec{r}%
-2(V+M)g(r)\text{ }\vec{r}\right] _{+}  \notag \\
&&+(1+\beta )[V,\vec{\Lambda}]+(1+\beta )\left[ p^{2},g(r)\text{ }\vec{r}%
\right]   \label{016}
\end{eqnarray}

The condition (23) gives%
\begin{equation}
f(r)=2(V+m)g(r)  \label{017}
\end{equation}%
and
\begin{equation}
(1+\beta )[V,\vec{\Lambda}]+(1+\beta )\left[ p^{2},g(r)\text{ }\vec{r}\right]
=0  \label{018}
\end{equation}

Now using Eq. (15) and the fact that $(1+\beta )(1-\beta )=0,$%
\begin{eqnarray}
(1+\beta )[V,\vec{\Lambda}] &=&(1+\beta )\left[ V,\vec{p}\times \vec{%
l}-\vec{l}\times \vec{p}\right]   \notag \\
&=&(1+\beta )\frac{i}{r}\frac{\partial V}{\partial r}\left[ \vec{r}%
\times \vec{l}-\vec{l}\times \vec{r}\right]   \label{019}
\end{eqnarray}%
On the other hand%
\begin{eqnarray}
\left[ p^{2},g(r)\text{ }\vec{r}\right]  &=&-2i\left[ g(r)+\frac{\partial g}{%
\partial r}\right] \vec{p}-\frac{i}{r}\frac{\partial g}{\partial r}\left[
\vec{r}\times \vec{l}-\vec{l}\times\vec{r}\right]  -\frac{1}{r}\left[ \frac{\partial }{\partial r}(g(r)+r\frac{\partial g}{%
\partial r})\right] \vec{r}  \label{020}
\end{eqnarray}%
It is important to point out that so far we have made no commitment
to the form of potentials $V(r)$ and $g(r).$ The condition (26) is
satisfied if
\begin{equation*}
g(r)=V(r)  \label{021}
\end{equation*}%
and%
\begin{equation}
V(r)+r\frac{\partial V}{\partial r}=0
\end{equation}%
which gives $\ V(r)=$constt. $\frac{1}{r}$ i.e, the Coulomb potential $-%
\frac{\alpha}{r}.$ It is the constraint (26) which forces the
relations (29) and as a result $V(r)$ has to be the Coulomb
potential. Thus
\begin{equation*}
2m\vec{\Gamma}=2(1+\beta )V(V+m)\text{ }\vec{r}+\vec{\Lambda}+\vec{F}
\end{equation*}%
where
\begin{equation}
\vec{F}=\gamma _{5}\left[ (1-\beta )V\text{ }\vec{r}\text{ }\vec{\Sigma}%
\cdot \vec{p}+h.c.\right]   \label{022}
\end{equation}%
In order to find $\frac{1}{4m^{2}}\left[ \Gamma ^{i},\Gamma ^{j}\right] $,
we note that%
\begin{equation}
\left[ \Lambda ^{i},\Lambda ^{j}\right] =-4\epsilon ^{ijk}\text{ }p^{2}L^{k}
\label{023}
\end{equation}%
\begin{equation}
\left[ (1+\beta )^{2}V(V+m)x^{i},\Lambda ^{j}\right] -i\leftrightarrow
j=-4(1+\beta )V(V+2M)i\epsilon ^{ijk}L^{k}
\end{equation}%
where we have used, $r\frac{\partial V}{\partial r}=-V(r)$ for $V=-\frac{\alpha}{r}$ and that [c.f. Eq. (\ref{010})]%
\begin{equation*}
(1+\beta )l^{k}=(1+\beta )L^{k}
\end{equation*}%
Further%
\begin{equation}
\left[ \Lambda ^{i},F^{j}\right] -i\leftrightarrow j=-4i\epsilon ^{ijk}\text{ }\gamma ^{5}\left[ (1-\beta )V\text{ }\vec{\Sigma}%
\cdot \vec{p}+(1+\beta )\text{ }\vec{\Sigma}\cdot \vec{p}V\right]L^{k}
\label{025}
\end{equation}%
\begin{equation}
\left[ (1+\beta )V(V+m)x^{i},F^{j}\right] -i\leftrightarrow j=0  \label{026}
\end{equation}%
\begin{equation}
\left[ F^{i},F^{j}\right] =-4(1+\beta )\epsilon^{ijk}L^{k}V^{2}
\label{027}
\end{equation}%
Collecting the various terms and using Eq. (\ref{012b}) with $V=-\alpha/r$we see that%
\begin{equation}
\left[ \Gamma ^{i},\Gamma ^{j}\right] =-\frac{H^{2}-m^{2}}{m^{2}}i\epsilon
^{ijk}L^{k}  \label{028}
\end{equation}%
Defining $K^{i}=\sqrt{\frac{m^{2}}{m^{2}-H^{2}}}\Gamma ^{i}$, we have
finally,%
\begin{eqnarray}
\left[ K^{i},K^{j}\right]  &=&i\epsilon ^{ijk}L^{k}  \notag
\\
\left[ K^{i},L^{j}\right]  &=&i\epsilon ^{ijk}K^{k}  \notag \\
\left[ L^{i},L^{j}\right]  &=&i\epsilon ^{ijk}L^{k}
\label{029}
\end{eqnarray}%
which generate $SO(4)$ algebra. Further $S^{i}$ commutes with
$L^{j}$ as well as with $H$ and $K^{j}$. Thus defining
\begin{eqnarray*}
M^{i} &=&\frac{L^{i}+K^{i}}{2} \\
N^{i} &=&\frac{L^{i}-K^{i}}{2}
\end{eqnarray*}%
we see that
\begin{eqnarray}
\left[ M^{i},M^{j}\right]  &=&i\epsilon ^{ijk}M^{k}  \notag\\
\left[ N^{i},N^{j}\right]  &=&i\epsilon ^{ijk}K^{k}  \notag \\
\left[ M^{i},N^{j}\right]  &=&0
\label{029a}
\end{eqnarray}
Thus the invariance group for the Dirac Hamiltonian (13) for the
Coulomb potential of the hydrogen atom is $SU_{M}\otimes
SU_{N}\otimes SU_{\sigma }(2)$, where $SU_{\sigma }(2)$ is the group
generated by $S^{i}$ given in Eq. (12).

The energy spectrum can now be easily determined
\begin{equation}
 \Gamma^{2}=\frac{H^{2}-m^{2}}{m^{2}}\left(L^{2}+1\right) +\alpha^{2}\frac{(H+m)^{2}}{m^{2}}\label{32a}
\end{equation}
and $\vec{\Gamma}\cdot\vec{L}=\vec{L}\cdot\vec{\Gamma}=0$, implying
$\vec{K}\cdot\vec{L}=0$ so that $M^{2}=N^{2}$. In terms of Casimir
operator, $M^{2}$, we can write Eq. $\eqref{32a}$ as
\begin{equation}
4(m^2-H^2)\vec{M}^{2}=H^{2}-m^{2} +\alpha^{2}(H+m)^{2}\label{35a}
\end{equation}
Now since $\vec{M}$ obey angular momentum commutation relations,
$\vec{M}^2$ has eigenvalues $\frak{m}(\frak{m}+1)$, where $\frak{m}$
can take on the values 0, 1/2, 1, $\cdots$. It is customary to use
$j$ for $\frak m$, then Eq. $\eqref{35a}$, gives the energy
eigenvalues
\begin{equation}
 E= m \frac{4 n^2-\alpha^2}{4n^2+\alpha^2} \label{36a}
\end{equation}
where $n=2j+1$, i.e. the energy spectrum is determined only by the
principal quantum number $n$ and states in the different $j$ values
are degenerate showing no spin-orbit splitting. In fact Eq. (41)
gives for $\epsilon=E-m$

\begin{equation*}
\epsilon_{n}=-m\frac{\alpha^{2}}{2n}+O(\alpha^{4})
\end{equation*}
which agrees with the energy spectrum for hydrogen atom in
Schrodinger theory.

The Dirac equation with vector and/or scalar Coulomb like potentials
\begin{eqnarray}
 V_{V}(r)&=&-\frac{\alpha_{V}}{r} \nonumber \\
V_{S}(r)&=&-\frac{\alpha_{S}}{r} \label{42}
\end{eqnarray}
is exactly solvable \cite{dt1} and energy spectrum is given by
\begin{equation}
E=\frac{m}{\alpha_{V}^{2}+(n-\delta_{j})^{2}}\left\{-\alpha_{V}\alpha_{S}\pm
(n-\delta_{j})\left[\alpha_{V}^{2}-\alpha_{S}^{2}+(n-\delta_{j})^{2}\right]^{1/2}\right\}\label{37a}
\end{equation}
where
\begin{equation}
 \delta_{j}=(j+\frac{1}{2})-\left[(j+\frac{1}{2})^{2}-(\alpha_{V}^{2}-\alpha_{S}^{2})\right]^{1/2}
\end{equation}
For $\alpha_{V}=\alpha_{S}$, this reduces to Eq. (\ref{36a}).

\section{$SU(3)$ symmetry for the Relativistic Harmonic Oscillator}

"The career of a young theoretical physicist consists of treating
the harmonic oscillator in ever-increasing levels of abstraction."
-- Sidney Coleman

In this section we discuss how harmonic oscillator provides a simple
realistic model to introduce $SU(3)$ symmetry \cite{moment} in a
language which is more familiar.

In non-relativistic quantum mechanics, the harmonic oscillator Hamiltonian%
\begin{equation}
H=\frac{1}{2m}[p^{2}+m^{2}\omega ^{2}r^{2}],  \label{30}
\end{equation}%
which is symmetric in $p\longleftrightarrow x$, commutes with the quadrupole
moment operator $Q^{ij},[i,j=1,2,3]$%
\begin{equation}
Q^{ij}=\left[
 m^{2}\omega ^{2}\left( x^{i}x^{j}-\frac{1}{3}\delta
^{ij}x^{2}\right) +\left( p^{i}p^{j}-\frac{1}{3}\delta ^{ij}p^{2}\right) %
\right]   \label{31}
\end{equation}%
Note that, being symmetric in $i$ and $j$ as well as traceless, $Q^{ij}$ has
five independent components while the orbital angular momentum $%
l^{ij}=x^{i}p^{j}-x^{j}p^{i}$, being antisymmetric in $i$ and $j$, has three
independent components. In order to go from tensor basis to the Gell-mann
basis, we introduce%
\begin{equation}
F_{a}=\left[ m^{2}\omega^{2}\overline{x}^{T}\frac{\lambda
_{a}}{2}\overline{x}+\frac{i}{2}m\omega\left( \overline{x}^{T}\frac{\lambda _{a}}{2}\overline{p}-\overline{p}^{T}%
\frac{\lambda _{a}}{2}\overline{x}\right) +\overline{p}^{T}\frac{\lambda _{a}%
}{2}\overline{p}\right]   \label{32}
\end{equation}%
where $a=1,....8,$ and $\overline{x}$ and $\overline{p}$\ are column
matrices [belonging to representation $3$ of $SU(3)$]%
\begin{equation}
\overline{x}=\left(
\begin{array}{c}
x^{1} \\
x^{2} \\
x^{3}%
\end{array}%
\right) ,\overline{p}=\left(
\begin{array}{c}
p^{1} \\
p^{2} \\
p^{3}%
\end{array}%
\right)   \label{33}
\end{equation}%
The superscript $T$ denotes transpose so that $\overline{x}^{T}$ and $%
\overline{p}^{T}$\ are row matrices. The tensor and Gell-Mann bases
are
related by%
\begin{eqnarray}
F_{a} &=&\frac{1}{2}\underset{i,j}{\sum }\left( \lambda _{a}\right)
_{ij}F^{ij}  \notag \\
F^{ij} &=&\underset{a}{\sum }\left( \lambda _{a}\right) ^{ij}F_{a}
\label{34}
\end{eqnarray}%
$\lambda _{a}$ are $3\times 3$\ Gell-Mann matrices [6]. The
decomposition (\ref{32}) corresponds to $3\times 3=3+\overline{6}$
where the representation $3$ is antisymmetric and corresponds to
orbital angular momentum $l^{ij}$ while the representation
$\overline{6}$\ is symmetric and correspond to the
quadropole moment $Q^{ij}$.\ In particular $%
F_{2}=m\omega l^{3},F_{5}=-m\omega l^{2},F_{7}=m\omega l^{1}$ while $F_{1},F_{3},F_{4},F_{6}$ and $%
F_{8}$\ correspond to $Q^{ij}$\ e.g. $%
F_{1}=m^2\omega^2 x^{1}x^{2}+p^{1}p^{2}=Q_{x}^{12}+Q_{p}^{12}$. Using the
commutation relations, [$I$ is $3\times 3$ unit matrix]
\begin{eqnarray}
\lbrack \overline{x},\overline{p}^{T}] &=&iI  \label{35} \\
\lbrack \frac{\lambda _{a}}{2},\frac{\lambda _{b}}{2}] &=&if_{abc}\frac{%
\lambda _{c}}{2}  \label{37}
\end{eqnarray}%
we obtain%
\begin{equation}
\lbrack F_{a},F_{b}]=if_{abc}m\omega F_{c}
\end{equation}%
where $a,b,c=2,5,7$\ and this corresponds to
\begin{equation}
\lbrack l^{i},l^{j}]=i\epsilon ^{ijk}l^{k}  \label{38}
\end{equation}%
On the other hand for $a,b=1,3,4,6,8$ in Eq. (\ref{37}), $c$ is restricted to $%
2,5,7$ and $%
f_{123}=1,(f_{147},f_{246},f_{257},f_{345})=1/2,(f_{156},f_{367})=-1/2,(f_{458},f_{678})=%
\sqrt{3}/2$, all others are zero.

Now for the relativistic harmonic oscillator where we have the Dirac
Hamiltonian (3), with $V(r)=\frac{1}{2}m\omega ^{2}r^{2}$,
qaudrupole moment operator takes the form%
\begin{eqnarray}
\Gamma ^{ij} &=&(1+\beta )f(r)Q_{x}^{ij}+Q_{p}^{ij} +\gamma
^{5}\left[ (1-\beta )g(r)Q_{x}^{ij}\vec{\Sigma }\cdot
\vec{p}+(1+\beta )\vec{\Sigma }\cdot\vec{p}%
g(r)Q_{x}^{ij}\right]   \label{39}
\end{eqnarray}%
Since $[\gamma ^{5}\vec{\Sigma
}\cdot\vec{p},Q_{p}^{ij}]=0$,%
\begin{eqnarray}
\lbrack H,\Gamma ^{ij}] &=&\gamma ^{5}[\vec{\Sigma }\cdot
\vec{p},\text{ }f(r)Q_{x}^{ij}-2(V+m)g(r)Q_{x}^{ij}]  +\gamma ^{5}\beta \lbrack \vec{\Sigma }\cdot\vec{p},%
\text{ }f(r)Q_{x}^{ij}-2(V+m)g(r)Q_{x}^{ij}]_{+}  \notag \\
&&+(1+\beta )[V,\text{ }Q_{p}^{ij}]+(1+\beta )[p^{2},\text{ }g(r)Q_{x}^{ij}]
\label{40}
\end{eqnarray}

Thus $[H,\Gamma ^{ij}]=0$ gives [with $Q_{x}^{ij}=m^{2}\omega
^{2}\left( x^{i}x^{j}-\frac{1}{3}\delta ^{ij}x^{2}\right) $ and
similar expression for $
Q_{p}^{ij}$]%
\begin{eqnarray}
f(r) &=&2(V+m)g(r)  \notag \\
g(r) &=&\frac{1}{2m}  \label{41}
\end{eqnarray}%
with $V(r)=\frac{1}{2}M\omega ^{2}r^{2}$. Thus%
\begin{eqnarray}
\Gamma ^{ij} &=&(1+\beta )\frac{1}{m}(V+m)Q_{x}^{ij}+Q_{p}^{ij}
+\frac{1}{2m}\gamma ^{5}\left[ (1-\beta )Q_{x}^{ij}\vec{\Sigma
}\cdot\vec{p}+(1+\beta )\vec{\Sigma }\cdot\vec{p} Q_{x}^{ij}\right]
\label{42}
\end{eqnarray}%
In the Gell-Mann basis this becomes
\begin{eqnarray}
\Gamma _{a} &=&(1+\beta )\frac{1}{m}(V+m)F_{a}^{x}+F_{a}^{p}
+\frac{1}{2m}\gamma ^{5}\left[ (1-\beta )F_{a}^{x} \vec{\Sigma
}\cdot\vec{p}+(1+\beta )\vec{\Sigma }\cdot\vec{p}F_{a}^{x}\right]
\label{43}
\end{eqnarray}%
Then using the commutation relations (50) and (51)
\begin{equation}
\lbrack \Gamma _{a},\Gamma _{b}]=if_{abc}\omega \left( (1+\beta
)(V+m)F_{c}+ \frac{1}{2}\gamma ^{5}\left[ (1-\beta )F_{c}\vec{\Sigma
}\cdot
\vec{p}+(1+\beta )\vec{\Sigma }\cdot\vec{p}%
F_{c}\right]
\right)   \label{44}
\end{equation}%
where $F_{c}=im\omega [\overline{x}^{T}\frac{\lambda _{c}}{2}\overline{p}-%
\overline{p}^{T}\frac{\lambda _{c}}{2}\overline{x}]$. Thus%
\begin{equation}
\lbrack \Gamma _{a},\Gamma _{b}]=if_{abc}\omega \left[ (H+m)F_{c}+\frac{%
1}{2}\gamma ^{5}(1-\beta )[F_{c},\vec{\Sigma }\cdot \vec{p}]\right]
\label{45}
\end{equation}%
where $a,b=1,3,4,6,8$ while $c=2,5,7$, so that $F_{c}$ is
essentially the orbital angular momentum $\vec{l}$. Now using Eqs.
(15) and (18)
\begin{eqnarray}
(H+m)L^{k} &=&\left[ \gamma ^{5}\vec{\Sigma }\cdot\vec{p}%
+(1+\beta )(V+m)\right] \left[ l^{k}+\frac{1}{2}(1-\beta )(\Sigma ^{k}-\frac{%
\vec{\Sigma }\cdot\vec{p}}{p^{2}}p^{k}\right]   \notag \\
&=&(H+m)l^{k}+\frac{i}{2}\gamma ^{5}(1-\beta )\vec{(\Sigma }%
\times \vec{p})^{k}  \notag \\
&=&(H+m)l^{k}+\frac{1}{2}\gamma ^{5}(1-\beta )\frac{1}{m\omega}\left[ F_{c},\vec{%
\Sigma }\cdot\vec{p}\right]   \label{46}
\end{eqnarray}%
where we have used $\frac{1}{m\omega}\left[ F_{c},\vec{\Sigma }\cdot\vec{p}%
\right] \sim \left[ l^{k},\vec{\Sigma }\cdot\vec{p}\right]
=\frac{i}{2}(\vec{\Sigma }\times \vec{p})^{k},$ Thus from Eqs. (60)
and (61)
\begin{equation}
\left[ \widetilde{\Gamma }_{a},\widetilde{\Gamma }_{b}\right] =if_{abc}%
\widetilde{F}_{c}  \label{47}
\end{equation}%
where now $\widetilde{F}_{2}=L^{3},\widetilde{F}_{5}=-L^{2},\widetilde{F}%
_{7}=L^{1}$ and $\widetilde{\Gamma }_{a}=\frac{\Gamma _{a}}{\sqrt{m\omega
^{2}(H+m)}}$. Further%
\begin{equation}
\left[ \widetilde{F}_{a},\widetilde{F}_{b}\right] =if_{abc}\widetilde{F}_{c}
\label{48}
\end{equation}%
$a,b,c=2,5,7$ and%
\begin{equation}
\left[ \widetilde{F}_{a},\widetilde{\Gamma }_{b}\right] =if_{abc}\widetilde{%
\Gamma }_{c}  \label{49}
\end{equation}%
where $a=2,5,7$ and $b,c=1,3,4,6,8$.

The commutation relations (\ref{47}, \ref{48}, \ref{49}) generate
the $SU(3)$ algebra. Since $\vec{S}$ commutes with $H$ as well as
with all the above generators, the invariant algebra of the
relativistic harmonic
oscillator represented by the Dirac Hamiltonian (3) with $V(r)=\frac{%
1}{2}m\omega ^{2}r^{2}$ is $SU(3)\otimes SU_{\sigma }(2)$.

We now calculate the energy spectrum for which purpose we note from
Eq. (58) that $[a=1,3,4,6,8]$
\begin{equation}
\Gamma^{2}=\sum_{a}\Gamma_{a}\Gamma_{a}=\frac{1}{3}\left(H^2-m^2\right)^{2}-m\omega^{2}(H+m)(L^2+3)\label{50}
\end{equation}
where we have used the definitions of $F_{a}^{x}$ and $F_{a}^{p}$
given in Eq. \eqref{30}, e.g. $F_{1}^{x}=m\omega^{2}x_{1}x_{2}$ and
$F_{1}^{p}=p_{1}p_{2}$, Eqs. (11) and (19) with
$V=\frac{1}{2}m\omega^{2}r^2$, $(1+\beta)(1-\beta)=0$,
$\gamma_{5}(1\pm\beta)=(1\mp\beta)\gamma_{5}$ and Eq. (15). In terms
of $\widetilde{\Gamma}_{a}$, Eq. \eqref{50} takes the form
\begin{equation}
3m\omega^{2}\left[\sum_{a}\widetilde{\Gamma}_{a}\widetilde{\Gamma}_{a}+L^{2}\right]=(H-m)^{2}(H+m)-9m\omega^{2}
\label{51}
\end{equation}
But $L^{2}=\sum_{b}\widetilde{F}_{b}^{2}$, where $b=2,5,7$. Thus Eq.
\eqref{51} takes the form
\begin{equation}
3m\omega^{2}\widetilde{\Gamma}^{2}=(H-m)^{2}(H+m)-9m\omega^{2}\label{52}
\end{equation}
where
$\widetilde{\Gamma}^{2}=\sum_{a}\widetilde{\Gamma}_{a}\widetilde{\Gamma}_{a}+\sum_{b}\widetilde{\Gamma}_{b}\widetilde{\Gamma}_{b}$
is the invariant of the group SU(3) and as such is proportional to
unit matrix.

Hence Eq. \eqref{52} gives the energy eigenvalues
\begin{equation}
(E-m)^{2}(E+m)-9m\omega^{2}=Cm\omega^{2}\label{53}
\end{equation}
where $C$ is to be fixed. This can be done in the following way. We
take the non-relativistic limit of Eq. (18), $H\to H_{non-rel}+m$,
which gives
\begin{equation}
H_{non-rel}=\frac{p^2}{2m}+2V=\frac{p^2}{2m}+m\omega^2 r^2\label{54}
\end{equation}
which, as is well known, gives the energy eigenvalues [note we have
to replace $\omega$ by $\sqrt{2}\omega$ in the ordinary harmonic
oscillator eigenvalues]
\begin{equation}
\epsilon_{N}=\frac{1}{\sqrt{2}}\omega (2N+3)\label{55}
\end{equation}
Now we take the non-relativistic limit of Eq. \eqref{53}, $E\to
{\cal E}_{N}+m$, which gives
\begin{equation}
m\omega^{2}(2N+3)^2-9m\omega^{2}=Cm\omega^{2}\label{56}
\end{equation}
fixing $C=4N(N+3)$. Putting back in Eq. (\ref{53}) the energy
eigenvalue equation becomes
\begin{equation}
(E_{N}-m)^{2}(E_{N}+m)=4\left(N+\frac{3}{2}\right)^{2}m\omega^{2}\label{57}
\end{equation}
where $N=0,1,\cdots$. This agrees with one obtained from the exact
solutions of Dirac equation \cite{12}.

\section{Summary and Conclusions}

\bigskip We have systematically reviewed the various dynamical symmetries of
the Dirac Hamiltonian, clearly stating the conditions under which
such symmetries hold. These symmetries include relativistic spin
(pseudo-spin) and orbital angular symmetries which hold when Dirac
Hamiltonian with scalar $V_{S}(r)$\ and vector $V_{V}(r)$
spherically symmetric potentials satisfy $\partial V_{V}/\partial
r=\pm \partial V_{S}/\partial r$ or $V_{V}=\pm V_{S}+U.$\ Here $U$
is a constant potential but can be absorbed in redefinition of the
mass.
Then if $V(r)=-\alpha/r,$ as for the hydrogen atom or in QCD the Dirac Hamiltonian has $%
SO(4)\otimes SU_{\sigma }(2)$ or equivalently $SU_{M}(2)\otimes
SU_{N}(2)\otimes SU_{\sigma }(2)$ symmetry. Here $SU_{\sigma }(2)$ is the $%
SU(2)$ group generated by the relativistic spin $S^{i}$ defined in
Eq. (8) or Eq. (12). If $V(r)=\frac{1}{2}M\omega ^{2}r^{2}$, the
harmonic oscillator potential, then the symmetry is $SU(3)\otimes
SU_{\sigma }(2)$. We have used Dirac algebra of Dirac matrices and
for the simple harmonic oscillation, Gell-Mann basis of $SU(3)$,
which is more transparent and simple (at least for physicist with
particle physics background) compared to the basis used in
\cite{03}. We have also calculated energy spectrum in each case from
group theoretical consideration, which agrees with the exact
solution of Dirac equation in each case. We have also indicated
physical systems where dynamical symmetries discussed above are
possibly relevant.


\begin{thebibliography}{99}

\bibitem{01} G. Baym, Lectures on Quantum Mechanics, The Benjamin/Cremmings
Puli. Com. Inc, (1969).

\bibitem{moment} D. M. Fradkin, Three-Dimensional Isotropic Harmonic Oscillator and $SU(3)$, Am. J. Phys. 33,
207-211 (1965).

\bibitem{02} J. P. Elliott, Collective Motion in the Nuclear Shell Model. I. Classification Schemes for States of Mixed Configurations, Proc. R.
Soc. \textbf{A 245}, 128-145 (1958).

\bibitem{03} J. N. Ginocchio, $U(3)$ and Pseudo-$U(3)$ Symmetry of the Relativistic Harmonic Oscillator, Phys. Rev. Letters, \textbf{95}, 252501-1--252501-3 (2005).

\bibitem{04} F. L. Zhang, B. Fu and J. L. Chen, Dynamical symmetry of Dirac hydrogen atom with spin symmetry
 and its connection with Ginocchio's oscillator, Phys. Rev. A \textbf{78},
040101-1--040101-4 (R) (2008).

\bibitem{05} See for example, Fayyazuddin and Riazuddin, A modern
introduction to Particle Physics, 2nd Edition, World Scientific
(2000).

\bibitem{06} See, Claude Itzykson and Jean-Bernard Zuber, Quantum Field Theory, McGraw-Hill International Editions
(1980).

\bibitem{dt1} W. Greiner, Relativistic Quantum Mechanics, Springer,
Berlin (2000); See also Symmetry of Dirac Equation and corresponding
phenomenology, Hong-Wei Ke et al. arxiv: 0907.0051, (2009).

\bibitem{dt2} J. S. Bell and H. Ruegg, Dirac equations with an exact higher symmetry, Nucl. Phys. B \textbf{98},
151-153 (1975); A. L. Blokhin, C. Bahri, and J. P. Draayer, Origin
of Pseudospin Symmetry, Phys. Rev. Lett. 74, 4149-4152 (1995); J. N.
Ginocchio and A. Leviatan, On the relativistic foundations of
pseudospin symmetry in nuclei, Phys. Lett. B \textbf{425}, 1-5
(1998); J. N. Ginocchio, A relativistic symmetry in nuclei, Phys.
Rep. \textbf{315}, 213-240 (1999).

\bibitem{07} A. Arima, M. Harrey, and K. Shimizu, Pseudo $LS$ coupling and pseudo $SU_{3}$ coupling schemes, Phys. Lett. B \textbf{30,}
517-522 (1969); K.Hecht and A. Adlen, Generalized seniority for
favored $J\neq0$ pairs in mixed configurations, Nucl. Physic. A 137,
129-143 (1969).

\bibitem{08} B. D. Serot and J. D. Walecka, in The Relativistic Nuclear
Many-Body Problem in Advances in Nuclear Physics, edited by J.W.
Negele and E. Vogt (Plenum, New York, 1986), Vol. 16; B. A.
Nikolaus, T. Hoch, and D. G. Madland, Nuclear ground state
properties in a relativistic point coupling model, Phys. Rev. C
\textbf{46}, 1757-1781 (1992).

\bibitem{010} K. Nakamura et al., Review of Particle Physics (Particle Data Group), J. Phys. G
\textbf{37}, 075021 (2010).

\bibitem{09} P. R. Page, T. Goldman and J. N. Ginocchio, Relativistic Symmetry Suppresses Quark Spin-Orbit Splitting, Phys. Rev. Lett.
\textbf{86}, 204-207 (2001).

\bibitem{12} J. N. Ginocchio, Relativistic harmonic oscillator with spin symmetry, Phys. Rev. C \textbf{69}, 034318-1--034318-8 (2004).

\end{thebibliography}
\end{document}